\documentclass[twocolumn,prl,aps]{revtex4}
\usepackage{graphicx}

\def\opone{\leavevmode\hbox{\small1\kern-3.8pt\normalsize1}}
\def\<{\langle}
\def\>{\rangle}

\def\opone{\leavevmode\hbox{\small1\kern-3.8pt\normalsize1}}
\newcommand{\beq}{\begin{equation}}
\newcommand{\eeq}{\end{equation}}
\newcommand{\beqa}{\begin{eqnarray}}
\newcommand{\eeqa}{\end{eqnarray}}

\begin{document}
\draft
\title{Quantum correlations versus Multisimultaneity: an experimental test}
\author{Andr\'{e} Stefanov, Hugo Zbinden, Nicolas Gisin}
\address{Group of Applied Physics, University of Geneva, 1211 Geneva 4, Switzerland}
\author{Antoine Suarez}
\address{Center for Quantum Philosophy, P.O. Box 304, CH-8044 Zurich}
\date{\today}

\begin{abstract}
Multisimultaneity is a causal model of relativistic quantum physics which
assigns a real time ordering to any set of events, much in the spirit of the
pilot-wave picture. Contrary to standard quantum mechanics, it predicts a
disappearance of the correlations in a Bell-type experiment when both
analysers are in relative motion such that, each one in its own inertial
reference frame, is first to select the output of the photons. We tested
this prediction using acousto-optic modulators as moving beam-splitters and
interferometers separated by 55 m. We didn't observe any disappearance of
the correlations, thus refuting Multisimultaneity.
\end{abstract}
\maketitle

Many experiments have demonstrated quantum correlations between spatially
separated measurements, under several conditions \cite{exp}, in perfect
concordance with quantum mechanical predictions. The most striking feature
of quantum correlations being the violation of Bell's inequalities \cite
{jb64}.

In this letter we confront quantum correlations with a natural alternative
model, called Multisimultaneity \cite{asvs97}. First, we summarize
Multisimultaneity, stressing its close relation to the famous pilot-wave
model of de Broglie and Bohm \cite{dbbh}. Next, we oppose the predictions of
quantum mechanics and of Multisimultaneity in the situation where two
entangled particles are analyzed by two beam-splitters moving apart in such
a way that each beam-splitter in its own inertial reference frame analyses
his particle before the other. We argue that Multisimultaneity is the
natural application of the pilot-wave intuition to this configuration. We
stress that it has the nice feature that it can be tested using available
technology. Finally, we present an experimental test based on 2-photon
interferences.

Within Newtonian physics, where time is absolute, it is possible to describe
quantum correlations at a distance in a causal ''mechanistic'' way \cite
{Eberhart}. A first class of examples assume that the collapse of the state
vector is a real physical phenomenon\cite{Pearle,GRW,PSD} : the first
measurement produces a collapse and the second measurement happens on a
system in the collapsed state.

Another explicit example, closer in spirit to the subject of this letter, is
provided by the pilot-wave model of de Broglie and Bohm. There, the particle
and the wave always co-exist, the wave guiding the particle and the particle
triggering the detectors. The two slit experiment is then not more difficult
to understand than the evolution of a cork floating in a river: if an island
separates the river in two arm over a certain length, then the cork passes
on one (and only one) side of the island, but its subsequent evolution is
also affected by the water that passes around the other side. When this
model is applied to two entangled particles, the model is less intuitive
(the ''wave flows'' in configuration space), but it still provides a clear
mechanistic description of how quantum correlations built up: the
measurement on one side modifies the ''wave'' which in turn guides the
distant particle (the model is local in configuration space, but nonlocal in
real space). This model reproduces all quantum predictions. If furthermore
one assumes that a privileges reference frame (e.g. defined by the cosmic
micro-wave background radiation) determines the time ordering, then this
model is self-consistent. However, when time is relative, as in special
relativity, it is ambiguous. Indeed, it is then no longer defined which
measurement modifies the wave first and which particle is then guided.

Multisimultaneity is an attempt to set the pilot-wave intuition in a
relativistic context. For this, one of the author of this letter, A. Suarez,
together with V. Scarani, proposed a model in which a causal temporal order
is defined for each measurement\cite{asvs97}. The basic idea is that the
relevant reference frame for each measurement is the inertial frame of the
massive apparatus. More specifically, Multisimultaneity assumes that the
relevant frame is determined by the analyzer's inertial frame (e.g. a
polarizer or a beam-splitter in our case). Paraphrasing Bohr, one could say
that the relevant frame, hence the relevant time ordering, depends on the
very condition of the experiment \cite{Bohr}. In Multisimultaneity, as in
the pilot-wave model, each particle emerging from a beam-splitter follows
one (and only one) outgoing mode, hence particles are always localized,
although the guiding wave (i.e. the usual quantum state $\psi $) follows all
path, in accordance with the usual Schr\"{o}dinger equation. When all
beam-splitters in an experiment are at relative rest, this model reduces to
the pilot-wave model and has thus precisely the same prediction as quantum
mechanics. However, when two beam-splitters move apart, then there are
several (i.e. two) relevant reference frames, each defining a time ordering,
hence the name of Multisimultaneity. In such a configuration it is possible
to arrange the experiment in such a way that each of the two beam-splitter
in its own reference frame analyses a particle from an entangled pair before
the other. In the pilot-wave picture, each particle has then to ''decide''
where to go before its twin particle makes its choice (even before the twin
is forced to make a choice). Multisimultaneity predicts that in such a {\it %
before-before} configuration, the correlations disappear, contrary to the
quantum prediction.

Let us emphasize that the model of Multisimultaneity, although conceptually
quite foreign both to quantum mechanics and to relativity, is not in
contradiction with any existing experimental data. Furthermore, it has the
nice feature that it can be tested using existing technology. In order to
realize a configuration with moving beam splitters, we proposed to use
traveling acoustic waves as beam splitter \cite{as00}.

Energy-time entanglement can be demonstrated by 2-photon interference
experiments\cite{WeihsTittel}. In this experiment (figure 1) we use the
Franson configuration \cite{Franson89}. Each photon from an energy-time
entangled photon pair source is sent to an analyzer. Each one consists in an
unbalanced interferometer, the difference between the long and short arms
being much longer than the coherence length of a single photon. The
coincidence events when both photons take the short arms or both the long
ones are indistinguishable because the emission time is undetermined, due to
the long coherence time of the pump laser. If we select only those events,
the coincidence rate between the 2 outputs is proportional to
\[
1+V\cos (\phi _{1}+\phi _{2})
\]
where $V$ is the visibility and $\phi _{1}$ and $\phi _{2}$ are the phase
differences between the long and the short arms in interferometers 1 and 2,
respectively. In an ideal case the visibility is $100\%$, but experimentally
it is always lower due, e.g., to detector noise and to partial
distinguishability of the interfering paths. Nevertheless, it can be larger
than 71\%, that is larger than the maximal visibility compatible with any
theory based only on local variable, as demonstrated by Bell's theorem \cite
{jb64}.

We use a recently developed PPLN waveguide source of energy-time entangled
photons\cite{SebElLett}. It features very high efficiency, so we can
register interferences in reasonable measuring times. Violation of Bell
inequality has already been demonstrated with this source \cite{PPLNQC}.

The beam-splitters have to be in motion, to achieve the desired relativistic
configuration. Therefore we built two unbalanced bulk Michelson
interferometers using AOM (Acousto-Optic Modulator, Brimrose
AMF-100-1.3-2mm) as beamsplitters. Indeed it would have been very difficult
to put real beamsplitters in motion at sufficiently high speed, but an AOM
can be seen as a realization of such a moving beamsplitter. Let us take a
look on the internal mechanism of the AOM: the traveling acoustic wave
inside the material changes the refractive index, thus creating a
diffraction grating. The reflection coefficient is maximal at Bragg angle $%
\theta _{B}$:
\begin{equation}
2\lambda _{s}\sin \theta _{B}=\lambda /n  \label{Bragg}
\end{equation}
where $\lambda _{s}$ is the sound wavelength, $\lambda $ is the light
wavelength in vacuum and $n$ the refractive index of the material. The
reflection coefficient is given by $R=\sin ^{2}\left( \sqrt{\alpha I}\right)
$ \cite{Yariv}, where $I$ is the acoustic power and $\alpha $ depends of the
AOM size and material, and of the light wavelength. Hence, the acoustic
power was set, such that the beamsplitting ratio is 50/50. The AOM ends by a
skew cut termination, thus the wave is damped rather than reflected, hence
the wave is effectively traveling rather than stationary. A point which
gives us confidence in using AOM \cite{as00} is that the reflection on a
moving mirror produces a frequency change of the light, due to the Doppler
effect, given by
\begin{equation}
\Delta \nu =\frac{2nv\sin \theta }{c}\nu  \label{Doppler}
\end{equation}
where $v$ is the mirror velocity and $\theta $ the angle between the
incident light and the plane of reflection. Within an AOM the reflected
light is also frequency shifted and the frequency shift is equal to the
acoustic wave frequency (100 MHz in our case):
\begin{equation}
\Delta \nu =\nu _{s}  \label{shift}
\end{equation}
Using $\lambda _{s}\nu _{s}=v_{s}$ for the sound wave and equations (\ref
{Bragg}) and (\ref{shift}) we found that the frequency shift induced by the
AOM is the same as the one induced by a mechanical grating traveling at
speed $v_{s}$.

The frequency shift induced by the AOMs on the reflected beam (200 MHz
because it passes twice through the modulator) forces us to use an\ AOM in
each interferometers, otherwise the total energy of both photons would not
be the same for the short-short and the long-long paths, leading to
distinguishability between the 2 paths. Therefore both interferometers are
build such that the shifts compensate, and the total energy remains the
same. Moreover both frequencies have to be exactly identical to avoid
beatings. This requires to synchronize the frequency of both AOM RF drivers
\cite{FullLength}.

Because of the small deviation angle (about 5$%
{{}^\circ}%
$) we collect only the light coming out from the input port by using an
fiber optical circulator (see figure 1). As the deviation angle depends of
the light wavelength, an AOM act as a band-pass filter for the reflected
beam with an bandwidth of about 30 nm. Therefore we have to insure that the
bandwidth of the photons is smaller by placing a spectral filter after the
source. The transmission through each interferometer is about 25\%.

When the phase $\phi $ is changed by slightly moving back and forth one of
the mirrors with a piezo-electric actuator, we observe 2-photon interference
fringes with a visibility of about 85\% after substraction of the accidental
coincidences (figure 2). The relatively low value for the visibility is
probably due to beatings originating from a small difference in the electric
spectrums of the RF signals driving the two AOMs.

In order to test Multisimultaneity, both interferometers have to be far
away, and both photons have to reach the moving beamsplitters at the same
time. The criterion given by special relativity for the change in time
ordering of two events in two reference frames counterpropagating at speed $%
v $ is
\begin{equation}
\left| \Delta t\right| <\frac{v}{c^{2}}d  \label{dt}
\end{equation}
where $\Delta t$ and $d$\ are respectively the time difference and distance
between the two events in the laboratory frame \cite{asvs97}. This criterion
is much more stringent than the spacelike separation condition $\left|
\Delta t\right| <\frac{d}{c}$. Due to the high speed of the acoustic wave
(2500 m/s, specified by the manufacturer or computed from the mechanical
properties of AMTIR (Amorphous Material Transmitting IR) \cite{handbook}), a
distance of 55 m between the interferometers is enough, and allows us to
realize the experiment inside our building. The permitted discrepancy on the
time of arrival of the photons in the AOM is then, according to (\ref{dt}), $%
1.53$ ps, corresponding to a distance of $0.46$ mm in air. The fiber path
length can me measured with a precision of 0.1 mm using a low coherence
interferometry method. The error on the interferometers' path lengths is
measured manually with a precision smaller than 0.5 mm. To be sure that we
have set the lengths as required we scan the path length difference by
pulling on a 1 m long fiber. The scan steps are of 0.12 mm. Simultaneously
we keep scanning the phase to observe interferences.

It is not sufficient to precisely equalize the path lengths, we also have to
insure that the coherence length of the photons is smaller than the
permitted discrepancy. This is the case because, with the interference
filter placed after the photon pair source, the photons coherence length is
about 0.14 mm. Another effect to take into account is the chromatic
dispersion in the fibres. This will spread the photon wave packet. However,
thanks to the energy correlation, the dispersion can be almost canceled. The
requirement for the 2-photon dispersion cancelation \cite{Steinberg92} is
that the center frequency of the 2 photon is equal to the zero dispersion
frequency of the fibers. We measured this value on a 2km fiber with a
commercial apparatus (EG\&G) which uses the phase shift method. We found a
value of $1313.2$ nm for $\lambda _{0}$. Then we used 100 m of the same
fiber assuming that the dispersion is equally distributed. We set the laser
wavelength at half this value. The pulse spreading over 100m, if we
conservatively assume a 1 nm difference between the laser wavelength and $%
\lambda _{0}$, is 0.2 ps \cite{Zbinden01}\ corresponding to a length of 0.06
mm in air, which is much smaller than the permitted discrepancy.

According to the Multisimultaneity model, the correlations should disappear
on a range of about 2-3 scanning steps. We scanned the path difference over
a range of $\pm $3 mm (figure 3), around the equilibrium point. Hence the
time order of the events passes from a before-after situation, where there
is a defined time ordering in all reference frames to the ambiguous
before-before condition and then back to an after-before situation. However
several scans like the one presented in figure 2 show no effects on the
visibility. A similar measure with both acoustic waves traveling against
each other ({\it after-after} condition \cite{asvs97}) doesn't show any
change of the correlations either. We can add that given the distance and
the incertitude on the time of arrival, we can fix a lower bound of the
speed in the laboratory frame of any hypothetical quantum influence to be $%
4.6\times 10^{5}c$.

Every day correlations are correlations between events; either the events
have a common cause or one event has a direct influence on the other(s).
That is, ordinary correlations is a secondary concept, built upon the
primary concept of event: the cause of ordinary correlations can be reduced
to the cause of the events. The violation of Bell's inequality rules out the
common cause explanation, and correlations between space-like separated
events exclude influences propagating slower than the speed of light.

Multisimultaneity is an alternative model to standard quantum mechanics in
which several reference frames, determined by the local physical devices and
apparatuses, each define a time-ordered causality with faster than light
influences [The influences being not under human control, they can't be used
for signaling]. In all situations where the different components of the
measuring apparatuses are at relative rest, Multisimultaneity has the same
prediction as quantum mechanics. However, in the intriguing situation where
entangled particles are analyses by two beam-splitters in relative motion
such that each one analyses ''his'' particle before the other,
Multisimultaneity predicts that the quantum correlations disappear. Since in
the reported experiment the correlations did not disappear,
Multisimultaneity is refuted.

This refutation stresses the oddness of quantum correlations. Not only are
they independent of the distance, but also it seems impossible to cast them
in any real time ordering. Recall that a model assuming that the detectors
determine the relevant frames has already been refuted \cite
{Zbinden01,Gisin00}. Hence one can't maintain any causal explanation in
which an earlier event influences a later one by arbitrarily fast
communication. In this sense, quantum correlations is a basic (i.e. primary)
concept, not a secondary concept reducible to that of causality between
events: Quantum correlations are directly caused by the quantum state in
such a way that one event cannot be considered the ``cause'' and the other
the ``effect''.

\section*{Acknowledgments}

This work would not have been possible without the financial support of the
''Fondation Odier de psycho-physique'' and the Swiss National Science
Foundation. We thank Valerio Scarani and Wolfgang Tittel for very
stimulating discussions and Fran\c{c}ois Cochet from Alcatel Cable Suisse SA
for the chromatic dispersion measurement apparatus.

\begin{figure}[h]
\includegraphics[width=8.43cm]{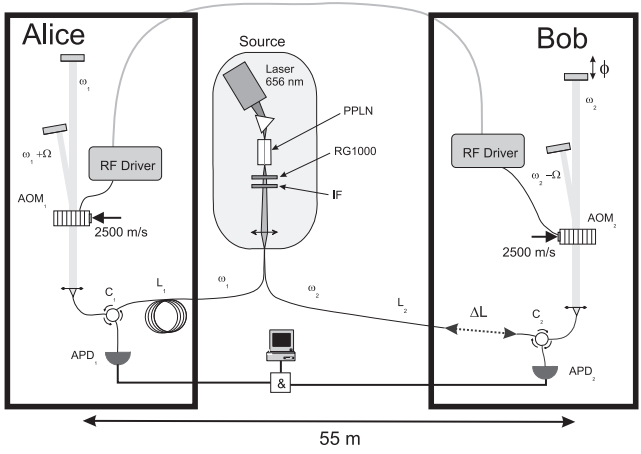}
\caption{Schematic of the experiment. The high efficiency photon pair source
uses a PPLN waveguide pumped by a 656 nm laser. RG1000 filter is used to
block the pump laser and a 11 nm large interference filter (IF) narrows the
photon bandwith. The two AOM's are 55 m apart and oriented such that the
acoustic wave propagate in opposite directions. One output of the
interferometers is collected back, thanks to optical circulators $C_{1}$ and
$C_{2}$ and the detection signals are send to a coincidence circuit. As the
frequency shifts are compensated, the total energy when both photons take
the short arms or the long ones is constant. 2-photons interference fringes
are observed by scanning the phase $\protect\phi$}
\end{figure}

\begin{figure}[h]
\includegraphics[width=8.43cm]{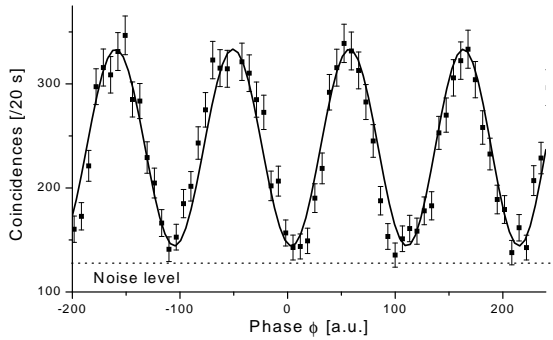}
\caption{Two-photons interference fringes. The sinusoidal fit gives a
visibility of 85$\pm $5\% after substraction of the accidental coincidences.}
\end{figure}

\begin{figure}[h]
\includegraphics[width=8.43cm]{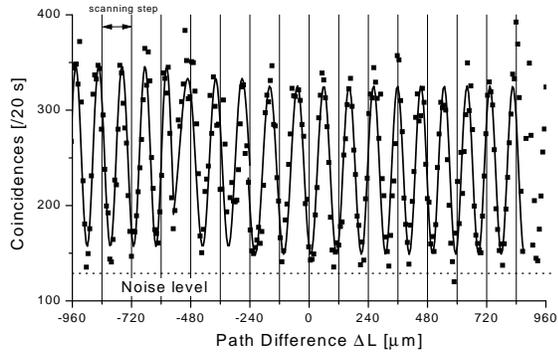}
\caption{Two-photon interference fringes. The total scan was done on a range
of $\pm $3 mm, but for more clarity the figure shows only a range of $\pm $1
mm around the estimated equilirium point. Each vertical line grid
corresponds to a a step of the scan. No disappearance of the correlation is
observed}
\end{figure}

\end{document}